\documentclass[fleqn,twoside,twocolumn,nofootinbib,showkeys]{revtex4} 
\usepackage[nocpr]{ujp} 

\usepackage{mathtext}           %
\usepackage{graphicx}
\usepackage{amsmath}
\usepackage{amssymb}
\usepackage{epsf}
\usepackage{calc}

\usepackage{pstricks-add,pst-coil}
\newlength{\mpicw}
\newlength{\mpich}
\newlength{\mpicth}
\newcounter{mpicth}
\newlength{\mpicwah}
\newlength{\mpicwbh}
\newcounter{mpicxa}
\newcounter{mpicxb}
\newcounter{mpicshift}
\newcounter{mpicfrom}
\newcounter{mpicto}
\newlength{\mpicya}
\newcounter{mpicya}
\newlength{\mpicyb}
\newcounter{mpicyb}
\newlength{\mpiclxab}

\newlength{\mpiclxac}
\newcounter{mpiclxac}
\newcounter{mpicly}

\begin{document}

\newcommand{\trule}[2]{%
\rule{#1em}{#2ex}%
}

\newcommand{\lrcoupling}[3]{%
\setlength{\mpicw}{\widthof{$#1#2#3$}}%
\setlength{\mpich}{\heightof{$#1#2#3$}}%
\setlength{\mpicth}{\mpich}%
\addtolength{\mpicth}{1.5ex}%
\setlength{\unitlength}{0.01\mpicw}%
\setcounter{mpicth}{100*\ratio{\mpicth}{\mpicw}}%
\setlength{\mpicwah}{\widthof{$#1$} * \real{0.5}}%
\setlength{\mpicwbh}{\widthof{$#3$} * \real{0.5}}%
\setlength{\mpicya}{\mpich}%
\addtolength{\mpicya}{0.33ex}%
\setcounter{mpicya}{100*\ratio{\mpicya}{\mpicw}}%
\setlength{\mpicyb}{\mpich}%
\addtolength{\mpicyb}{0.9ex}%
\setcounter{mpicyb}{100*\ratio{\mpicyb}{\mpicw}}%
\setcounter{mpicly}{\value{mpicyb}}%
\addtocounter{mpicly}{-\value{mpicya}}%
\setlength{\mpiclxab}{\mpicw}%
\addtolength{\mpiclxab}{-\mpicwah}%
\addtolength{\mpiclxab}{-\mpicwbh}%
\setcounter{mpicshift}{100*\ratio{2.5pt}{\mpicw}}%
\setcounter{mpiclxac}{50*\ratio{\mpiclxab}{\mpicw}}%
\setcounter{mpicxa}{100*\ratio{\mpicwah}{\mpicw}}%
\setcounter{mpicxb}{\value{mpicxa}}%
\addtocounter{mpicxb}{\value{mpiclxac}}%
\addtocounter{mpicxb}{\value{mpiclxac}}%
\setcounter{mpicfrom}{\value{mpiclxac}}%
\addtocounter{mpicfrom}{\value{mpicshift}}%
\setcounter{mpicto}{\value{mpiclxac}}%
\addtocounter{mpicto}{-\value{mpicshift}}%
\begin{picture}(100,\value{mpicth})%
\put(0,0){$#1#2#3$}%
\put(\value{mpicxa},\value{mpicya}){\line(0,1){\value{mpicly}}}%
\put(\value{mpicxa},\value{mpicyb}){\vector(1,0){\value{mpicfrom}}}%
\put(\value{mpicxb},\value{mpicyb}){\line(-1,0){\value{mpicto}}}%
\put(\value{mpicxb},\value{mpicya}){\line(0,1){\value{mpicly}}}%
\end{picture}%
}

\newcommand{\rlcoupling}[3]{%
\setlength{\mpicw}{\widthof{$#1#2#3$}}%
\setlength{\mpich}{\heightof{$#1#2#3$}}%
\setlength{\mpicth}{\mpich}%
\addtolength{\mpicth}{1.5ex}%
\setlength{\unitlength}{0.01\mpicw}%
\setcounter{mpicth}{100*\ratio{\mpicth}{\mpicw}}%
\setlength{\mpicwah}{\widthof{$#1$} * \real{0.5}}%
\setlength{\mpicwbh}{\widthof{$#3$} * \real{0.5}}%
\setlength{\mpicya}{\mpich}%
\addtolength{\mpicya}{0.33ex}%
\setcounter{mpicya}{100*\ratio{\mpicya}{\mpicw}}%
\setlength{\mpicyb}{\mpich}%
\addtolength{\mpicyb}{0.9ex}%
\setcounter{mpicyb}{100*\ratio{\mpicyb}{\mpicw}}%
\setcounter{mpicly}{\value{mpicyb}}%
\addtocounter{mpicly}{-\value{mpicya}}%
\setlength{\mpiclxab}{\mpicw}%
\addtolength{\mpiclxab}{-\mpicwah}%
\addtolength{\mpiclxab}{-\mpicwbh}%
\setcounter{mpicshift}{100*\ratio{2.5pt}{\mpicw}}%
\setcounter{mpiclxac}{50*\ratio{\mpiclxab}{\mpicw}}%
\setcounter{mpicxa}{100*\ratio{\mpicwah}{\mpicw}}%
\setcounter{mpicxb}{\value{mpicxa}}%
\addtocounter{mpicxb}{\value{mpiclxac}}%
\addtocounter{mpicxb}{\value{mpiclxac}}%
\setcounter{mpicfrom}{\value{mpiclxac}}%
\addtocounter{mpicfrom}{\value{mpicshift}}%
\setcounter{mpicto}{\value{mpiclxac}}%
\addtocounter{mpicto}{-\value{mpicshift}}%
\begin{picture}(100,\value{mpicth})%
\put(0,0){$#1#2#3$}%
\put(\value{mpicxa},\value{mpicya}){\line(0,1){\value{mpicly}}}%
\put(\value{mpicxa},\value{mpicyb}){\line(1,0){\value{mpicto}}}%
\put(\value{mpicxb},\value{mpicyb}){\vector(-1,0){\value{mpicfrom}}}%
\put(\value{mpicxb},\value{mpicya}){\line(0,1){\value{mpicly}}}%
\end{picture}%
}

\newcommand{\lrcouplingX}[5]{%
\setlength{\mpicw}{\widthof{$#1#2#3$}}%
\setlength{\mpich}{\heightof{$#1#2#3$}}%
\setlength{\mpicth}{\mpich}%
\addtolength{\mpicth}{1.5ex}%
\setlength{\unitlength}{0.01\mpicw}%
\setcounter{mpicth}{100*\ratio{\mpicth}{\mpicw}}%
\setlength{\mpicwah}{\widthof{$#1$} * \real{0.5}* \real{#4}}%
\setlength{\mpicwbh}{\widthof{$#3$} * \real{0.5}* \real{#5}}%
\setlength{\mpicya}{\mpich}%
\addtolength{\mpicya}{0.33ex}%
\setcounter{mpicya}{100*\ratio{\mpicya}{\mpicw}}%
\setlength{\mpicyb}{\mpich}%
\addtolength{\mpicyb}{0.9ex}%
\setcounter{mpicyb}{100*\ratio{\mpicyb}{\mpicw}}%
\setcounter{mpicly}{\value{mpicyb}}%
\addtocounter{mpicly}{-\value{mpicya}}%
\setlength{\mpiclxab}{\mpicw}%
\addtolength{\mpiclxab}{-\mpicwah}%
\addtolength{\mpiclxab}{-\mpicwbh}%
\setcounter{mpicshift}{100*\ratio{2.5pt}{\mpicw}}%
\setcounter{mpiclxac}{50*\ratio{\mpiclxab}{\mpicw}}%
\setcounter{mpicxa}{100*\ratio{\mpicwah}{\mpicw}}%
\setcounter{mpicxb}{\value{mpicxa}}%
\addtocounter{mpicxb}{\value{mpiclxac}}%
\addtocounter{mpicxb}{\value{mpiclxac}}%
\setcounter{mpicfrom}{\value{mpiclxac}}%
\addtocounter{mpicfrom}{\value{mpicshift}}%
\setcounter{mpicto}{\value{mpiclxac}}%
\addtocounter{mpicto}{-\value{mpicshift}}%
\begin{picture}(100,\value{mpicth})%
\put(0,0){$#1#2#3$}%
\put(\value{mpicxa},\value{mpicya}){\line(0,1){\value{mpicly}}}%
\put(\value{mpicxa},\value{mpicyb}){\vector(1,0){\value{mpicfrom}}}%
\put(\value{mpicxb},\value{mpicyb}){\line(-1,0){\value{mpicto}}}%
\put(\value{mpicxb},\value{mpicya}){\line(0,1){\value{mpicly}}}%
\end{picture}%
}

\newcommand{\rlcouplingX}[5]{%
\setlength{\mpicw}{\widthof{$#1#2#3$}}%
\setlength{\mpich}{\heightof{$#1#2#3$}}%
\setlength{\mpicth}{\mpich}%
\addtolength{\mpicth}{1.5ex}%
\setlength{\unitlength}{0.01\mpicw}%
\setcounter{mpicth}{100*\ratio{\mpicth}{\mpicw}}%
\setlength{\mpicwah}{\widthof{$#1$} * \real{0.5}* \real{#4}}%
\setlength{\mpicwbh}{\widthof{$#3$} * \real{0.5}* \real{#5}}%
\setlength{\mpicya}{\mpich}%
\addtolength{\mpicya}{0.33ex}%
\setcounter{mpicya}{100*\ratio{\mpicya}{\mpicw}}%
\setlength{\mpicyb}{\mpich}%
\addtolength{\mpicyb}{0.9ex}%
\setcounter{mpicyb}{100*\ratio{\mpicyb}{\mpicw}}%
\setcounter{mpicly}{\value{mpicyb}}%
\addtocounter{mpicly}{-\value{mpicya}}%
\setlength{\mpiclxab}{\mpicw}%
\addtolength{\mpiclxab}{-\mpicwah}%
\addtolength{\mpiclxab}{-\mpicwbh}%
\setcounter{mpicshift}{100*\ratio{2.5pt}{\mpicw}}%
\setcounter{mpiclxac}{50*\ratio{\mpiclxab}{\mpicw}}%
\setcounter{mpicxa}{100*\ratio{\mpicwah}{\mpicw}}%
\setcounter{mpicxb}{\value{mpicxa}}%
\addtocounter{mpicxb}{\value{mpiclxac}}%
\addtocounter{mpicxb}{\value{mpiclxac}}%
\setcounter{mpicfrom}{\value{mpiclxac}}%
\addtocounter{mpicfrom}{\value{mpicshift}}%
\setcounter{mpicto}{\value{mpiclxac}}%
\addtocounter{mpicto}{-\value{mpicshift}}%
\begin{picture}(100,\value{mpicth})%
\put(0,0){$#1#2#3$}%
\put(\value{mpicxa},\value{mpicya}){\line(0,1){\value{mpicly}}}%
\put(\value{mpicxa},\value{mpicyb}){\line(1,0){\value{mpicto}}}%
\put(\value{mpicxb},\value{mpicyb}){\vector(-1,0){\value{mpicfrom}}}%
\put(\value{mpicxb},\value{mpicya}){\line(0,1){\value{mpicly}}}%
\end{picture}%
}

\title[Energy Spectrum of the Pseudospin-Electron
Model]
{ENERGY SPECTRUM OF THE PSEUDOSPIN-ELECTRON MODEL IN A DYNAMICAL
MEAN-FIELD APPROACH}%
\author{I.V.~Stasyuk}
\affiliation{Institute for Condensed Matter Physics of the Nat. Acad. of
Sci. of Ukraine}
\address{1, Svientsitskii Str., Lviv 79011, Ukraine}

\author{V.O.~Krasnov}
\affiliation{Institute for Condensed Matter Physics of the Nat. Acad. of
Sci. of Ukraine}%
\address{1, Svientsitskii Str., Lviv 79011, Ukraine}%

\udk{538.945}  \razd{\secvii}


 \autorcol{I.V.\hspace*{0.7mm}Stasyuk, V.O.\hspace*{0.7mm}Krasnov}

\setcounter{page}{68}%

\begin{abstract}
The pseudospin-electron model in the case of infinite on-site electron
repulsion is investigated. The electron energy spectrum is
calculated within the framework of the dynamical mean field theory
(DMFT), and the alloy analogy approximation is developed. The effect of
the pseudospin-electron interaction, local asymmetry field, and
tunneling-like level splitting on the existence and the number of
electron subbands is investigated. The relation of
the pseudospin-electron model to the problem of energy spectrum of
boson-fermion mixtures in optical lattices is discussed.
\end{abstract}

\maketitle

\section{Introduction}
The pseudospin-electron model (PEM) is one of the models, which are used in
the physics of strongly correlated electron systems in recent years.
Application of the model to high-temperature superconductors allows one,
for example, to describe the thermodynamics of an anharmonic oxygen ion
subsystem and to explain the occurrence of inhomogeneous states and
the  bistability phenomena (\cite{Mull}, see also \cite{Book}).
In this model, one considers the dynamics of locally
anharmonic structure elements (using the pseudospin variables to
describe them), interaction between pseudospins and electrons, and
the asymmetry of local anharmonic potential wells. The electron
subsystem is described by the Hubbard Hamiltonian.

In \cite{Kra1}, a pseudospin-electron model of ion intercalation
in crystals was formulated; the pseudospin representation is also
used in the description of ionic conductors \cite{Mahan,Dulepa}.
The thermodynamics of the model has been investigated in the
mean-field approximation; it was shown that a new phase with
$\langle S^x \rangle \neq 0$ appears due to the ion hopping
between sites. This new phase is an analogue of the superfluid
phase in the systems of hard-core bosons or the superionic phase
in ionic (protonic) conductors.

The pseudospin-electron model in its general form has also a direct
relation to the ultra-cold boson-fermion mixtures in optical
lattices, which are the object of an intense theoretical
investigation during last years \cite{Albus,Lewen,IskFree,Mer,Heb}.
In the hard-core boson limit (that corresponds to the large on-site
repulsion between Bose atoms), when the pseudospin representation is
used, one has to do with a pseudospin-fermion analogue of PEM
\cite{taras}. The transverse field exists here in the phase with a
Bose--Einstein (BE) condensate and is proportional to the order
parameter $\langle S^x \rangle$, while the longitudinal field plays
the role of a chemical potential of bosons. In this case, the
investigation of the electron spectrum within PEM gives, at the same
time, an information about the fermion spectrum of the boson-fermion
system with $n_b \leq 1$ in the mentioned~phase.

Investigations of the thermodynamics and the dynamics of PEM were
performed mostly within the Hartree--Fock-type approximation (in the
weak pseudospin-electron coupling case) or the generalized random
phase approximation (at the strong coupling); the problem is
reviewed and discussed in \cite{Book}. The conditions of the
appearance of modulated phases, phase transitions between different
uniform phases, and phase separation were established.

In the DMFT approach, the Hamiltonian with strong correlations is taken
in the infinite space dimension ($d\rightarrow\infty$) limit; this
leads to a reformulation of the problem and the transition to the solution of
the single-site problem described by an effective Hamiltonian
\cite{Voll, Metz, Mull-Hart}. Only for simplest cases such as mobile
particles in the Falicov--Kimball model, one can solve analytically
this problem. An exact solution exists also for the pseudospin-electron
model without transverse field \cite{JPS}. There
are also some approximate analytical approaches such as: Hubbard-I,
Hubbard-III, alloy analogy (AA), modified alloy analogy (MAA) {\it
etc}., see \cite{Pott, Sta1}.

In this work, the alloy analogy approximation for the single-site
problem is used within the pseudospin-electron model. Our problem is solved
in the limit of infinite value of single-site electron interaction.
The previous consideration of this problem in the
Hubbard-I approximation \cite{Stash} revealed a complicated structure
of the spectrum and the presence of some number of subbands. In
\cite{My1}, the electron energy spectrum of the pseudospin-electron
model allowing the interaction of near-energy subbands was
considered. The effective single-site problem was solved within the
auxiliary fermion field approach with the help of the procedure of
different-time decoupling of higher order Green's functions
\cite{Sta1}.

Our main task is to investigate a reconstruction of the electron
energy spectrum and to describe the effect of band splitting at a
change of the longitudinal and transverse fields and the
pseudospin-electron interaction constant.

\section{Hamiltonian of the Model and its Transformation}

The Hamiltonian of the pseudospin-electron model is
\begin{equation}
 H = \sum\limits_{i}H_{i}+
\sum\limits_{<i,j>,\sigma} t_{ij}c^{+}_{i\sigma}c_{j\sigma} ,
\end{equation}
where $ \Omega $ in the single-site part of the Hamiltonian is the
tunneling-like level splitting, $g$ is the pseudospin-electron
interaction constant, and $h$ is the asymmetry of the local anharmonic
potential:
\begin{equation}
 H_{i}=-\mu(n_{i,\uparrow}+n_{i,\downarrow})+g(n_{i,\uparrow}+n_{i,\downarrow})S_{i}^{z}-
\Omega S_{i}^{x}-hS_{i}^{z} .
\end{equation}
The second term in (1) describes the electron site-to-site hopping
(between nearest neighbours).

The term $Un_{\uparrow}n_{\downarrow}$ is not included, as we
investigate our problem in the limit of the infinite potential of
single-site electron interaction $U\rightarrow \infty$, where all
states with double cite occupation are absent. The single-site
Hamiltonian is considered as the zero-order one with respect to
the electron transfer. It is useful to introduce the standard
single-site basis $
|R\rangle=|n_{i,\uparrow},n_{i,\downarrow},S_{i}^{z}\rangle $,
with six eigenvectors \cite{Stash}:
\[|1\rangle=\bigg|0,0,\frac{1}{2}\!\bigg\rangle,~~
|3\rangle=\bigg|0,1,\frac{1}{2}\!\bigg\rangle,~~
|4\rangle=\bigg|1,0,\frac{1}{2}\!\bigg\rangle,\]\vspace*{-7mm}
\begin{equation}
|\widetilde{1}\rangle=\bigg|0,0,-\frac{1}{2}\!\bigg\rangle,~~
|\widetilde{3}\rangle=\bigg|0,1,-\frac{1}{2}\!\bigg\rangle,~~
|\widetilde{4}\rangle=\bigg|1,0,-\frac{1}{2}\!\bigg\rangle.
\end{equation}

Using Hubbard $X$-operators that act in the space of such
eigenvectors, we can write down the electron annihilation (creation)
operators and the pseudospin operators (in the limit, when the state
$|2\rangle=|1,1,\pm \frac{1}{2}\rangle$ is unoccupied) as follows:
\cite{Stash}
\[c_{i,\uparrow}=\mathrm{X}_{i}^{14}+\mathrm{X}_{i}^{\widetilde{1}\widetilde{4}} ,\]\vspace*{-5mm}
\begin{equation}
c_{i,\downarrow}=\mathrm{X}_{i}^{13}+\mathrm{X}_{i}^{\widetilde{1}\widetilde{3}}
\end{equation}

Then,  the single-site part of the Hamiltonian can be expressed by means
of $X$-operators
in the following way:
\[H_{i}=(-\mu+\frac{g}{2}-\frac{h}{2})(\mathrm{X}_{i}^{33}+\mathrm{X}_{i}^{44})+\]\vspace*{-5mm}
\[+(-\mu-\frac{g}{2}+\frac{h}{2})(\mathrm{X}_{i}^{\widetilde{3}\widetilde{3}}+
\mathrm{X}_{i}^{\widetilde{4}\widetilde{4}})+\frac{h}{2}(\mathrm{X}_{i}^{\widetilde{1}\widetilde{1}}
-\mathrm{X}_{i}^{11})+\]\vspace*{-5mm}
\begin{equation}
+\frac{\Omega}{2}(\mathrm{X}_{i}^{1\widetilde{1}}+\mathrm{X}_{i}^{\widetilde{1}1}
+\mathrm{X}_{i}^{3\widetilde{3}}+\mathrm{X}_{i}^{\widetilde{3}3}+\mathrm{X}_{i}^{\widetilde{4}4}
+\mathrm{X}_{i}^{4\widetilde{4}}) .
\end{equation}
This Hamiltonian is diagonal in the case $\Omega=0$. But if the
tunneling splitting is non-zero, we have to use a transformation
\begin{eqnarray}
\binom{R}{\widetilde{R}}=
\left(\!\! \begin{array}{cc} \cos\phi_{r} & \sin\phi_{r} \\
-\sin\phi_{r} & \cos\phi_{r}
\end{array}\!\! \right)\binom{r}{\widetilde{r}}
\end{eqnarray}
to diagonalize it.
Here,
\[\cos(2\phi_{r})=\frac{n_{r}g-h}{\sqrt{(n_{r}g-h)^{2}+\Omega^{2}}}
,\]\vspace*{-5mm}
\begin{equation}
n_{1}=0,n_{3}=n_{4}=1.
\end{equation}
In that way, we have
\[H=\sum\limits_{i,r}\varepsilon_{r}X_{i}^{rr}+\sum\limits_{<i,j>,\sigma}t_{ij}c_{i\sigma}^{+}c_{j\sigma} ,\]\vspace*{-5mm}
\[\varepsilon_{1,\widetilde{1}}=\pm\frac{1}{2}\sqrt{h^{2}+\Omega^{2}},\]\vspace*{-5mm}
\begin{equation}
\varepsilon_{3,\widetilde{3}}=\varepsilon_{4,\widetilde{4}}=-\mu\pm\frac{1}{2}\sqrt{(g-h)^{2}+\Omega^{2}}.
\end{equation}
Here, making transformation (6), we obtain
\[c_{i,\uparrow}=\cos{\phi_{41}}(X_{i}^{14}+X_{i}^{\widetilde{1}\widetilde{4}})
+\sin{\phi_{41}}(X_{i}^{1\widetilde{4}}-X_{i}^{\widetilde{1}4}),\]\vspace*{-5mm}
\[c_{i,\downarrow}=\cos{\phi_{31}}(X_{i}^{13}+X_{i}^{\widetilde{1}\widetilde{3}})
+\sin{\phi_{31}}(X_{i}^{1\widetilde{3}}-X_{i}^{\widetilde{1}3}),\]\vspace*{-5mm}
\[\cos{\phi_{41}}=\cos{(\phi_4-\phi_1)}, \quad
\cos{\phi_{31}}=\cos{(\phi_3-\phi_1)},\]\vspace*{-5mm}
\begin{equation}
\sin{\phi_{41}}=\sin{(\phi_4-\phi_1)}, \quad
\sin{\phi_{31}}=\sin{(\phi_3-\phi_1)}.
\end{equation}
where $X$-operators act on the new basis.

\section{Dynamical Mean Field Theory Approach}

The transition to the $d=\infty$ limit in the DMFT approach is
accompanied by the scaling of the electron transfer parameter:
\begin{equation}
t_{ij} \rightarrow \frac{t_{ij}}{\sqrt{d}}.
\end{equation}
In particular, the self-energy part of electron Green's function
becomes purely local \cite{Metz,Mull-Hart}:
\begin{equation}
\Sigma_{ij,\sigma}(\omega)=\Sigma_{\sigma}\delta_{ij}, \qquad
d=\infty .
\end{equation}
The Fourier-transform $\Sigma_{ij,\sigma}(\omega)$ is, hence,
momentum-independent:
\begin{equation}
\Sigma_{\sigma}({\bf k},\omega)=\Sigma_{\sigma}(\omega).
\end{equation}
Electron Green's function in the $(k,\omega)$ representation
\begin{equation}
G_{k}^{\sigma}(\omega)=\sum\limits_{i-j}e^{i{\bf k}({\bf R}_{i}-{\bf
R}_{j})}G_{ij,\sigma}(\omega)
\end{equation}
can be expressed as
\begin{equation}
G_{k}^{\sigma}(\omega)=\frac{1}{[\Xi_{\sigma}(\omega)]^{-1}-t_{k}},
\end{equation}
where $ \Xi_{\sigma}(\omega)
$ is the part, which is irreducible (in the diagrammatic
representation) according to Larkin. To calculate the $
\Xi_{\sigma}(\omega)$ function, the effective single-site problem is
used. The transition to this problem corresponds to the replacement
\[e^{-\beta H}\to e^{-\beta
H_{\rm eff}}=e^{-\beta H_{0}}\,T {\rm
exp}\bigg\{\!\!-\int\limits_{0}^{\beta}d\tau\times\]\vspace*{-7mm}
\begin{equation}
\times\!\int\limits_{0}^{\beta}\!d\tau^{'}
\sum\limits_{\sigma}\!J_{\sigma}(\tau-\tau^{'})a_{\sigma}^{+}(\tau)a_{\sigma}(\tau^{'})\!\bigg\}\equiv
e^{-\beta H_0}\widetilde{\sigma}(\beta),
\end{equation}\vspace*{-7mm}

\noindent where
\begin{equation}
H_{0}=H_{i},
\end{equation}
and $J_{\sigma}(\tau-\tau^{'})$ is an effective time-dependent field
(coherent potential) that is determined self-consistently from the
condition that the same self-energy part $ \Xi_{\sigma}(\omega) $
determines the lattice function $ G_{k}^{\sigma}(\omega),$ as well
as the Green's function $G_{\sigma}^{(a)}(\omega)$ of the effective
single-site problem:
\begin{equation}
G_{\sigma}^{(a)}(\omega)=\frac{1}{[\Xi_{\sigma}(\omega)]^{-1}-J_{\sigma}(\omega)}.
\end{equation}
In this case, we have
\begin{equation}
G_{\sigma}^{(a)}=G_{ii,\sigma}(\omega)=\frac{1}{N}\sum\limits_{k}G_{k}^{\sigma}(\omega).
\end{equation}
The system of simultaneous equations (14), (17), and (18) becomes
closed, when it is supplemented with the functional dependence
\begin{equation}
G_{\sigma}^{(a)}(\omega)=f([J_{\sigma}(\omega)]),
\end{equation}
which is obtained as a result of solving the effective single-site
problem with the statistical operator $ \exp(-\beta H_{\rm eff})$.

\section{Reformulation of Wick's Theorem for~the~Single-Site Problem}

To find out relation (19), let us calculate electron
Green's function using an expansion in powers of the coherent potential
$J_{\sigma}(\omega)$. In the zero approximation:
\begin{equation}
-\langle T
X^{qp}(\tau)X^{pq}(\tau')\rangle_{0}=-g_{0}^{qp}(\tau-\tau')\langle
X^{qq}+X^{pp}\rangle_0.
\end{equation}
In the frequency representation: $g_0^{qp}(\omega_n)=-(i\omega_n-$
$-\lambda_{pq})^{-1}$, $\lambda_{pq}=\varepsilon_p-\varepsilon_q$.

Using Wick's theorem for the Hubbard operators \cite{Wick}, we can see that
\[\rlcoupling{X}{\vphantom{X}^{41}(\tau')}{X}
\vphantom{X}^{14}(\tau)=-g_{0}^{14}(\tau-\tau')(X^{11}+X^{44})_{\tau'}\]\vspace*{-7mm}
\[\rlcoupling{X}{\vphantom{X}^{\widetilde{4}\widetilde{1}}(\tau')}{X}
\vphantom{X}^{14}(\tau)=0\]\vspace*{-7mm}
\[\rlcoupling{X}{\vphantom{X}^{\widetilde{4}1}(\tau')}{X}
\vphantom{X}^{14}(\tau)=-g_{0}^{14}(\tau-\tau')X^{\widetilde{4}4}(\tau')\]\vspace*{-7mm}
\[\rlcoupling{X}{\vphantom{X}^{4\widetilde{1}}(\tau')}{X}
\vphantom{X}^{14}(\tau)=-g_{0}^{14}(\tau-\tau')X^{1\widetilde{1}}(\tau')\]
As a result of such a procedure, the Bose-type $X$-operators appear.

The alloy analogy approximation (see \cite{Pott, Sta1}) means
the neglect of all non-diagonal Hubbard operators in Wick's
pairings. Such an approximation leads to the next result:
\[\rlcoupling{c}{\vphantom{c}^{+}_{\uparrow}(\tau')}{X}\vphantom{X}^{14}(\tau)
=-g_{0}^{14}(\tau-\tau')(X^{11}+X^{44})_{\tau'}\cos{\phi_{41}},\]\vspace*{-7mm}
\[\rlcoupling{c}{\vphantom{c}^{+}_{\downarrow}(\tau')}{X}\vphantom{X}^{14}(\tau)=0,\]\vspace*{-7mm}
\[\rlcoupling{c}{\vphantom{c}^{+}_{\uparrow}(\tau')}{X}\vphantom{X}^{\widetilde{1}\widetilde{4}}(\tau)
=-g_{0}^{\widetilde{1}\widetilde{4}}(\tau-\tau')(X^{\widetilde{1}\widetilde{1}}
+X^{\widetilde{4}\widetilde{4}})_{\tau'}\cos{\phi_{41}},\]
\[\rlcoupling{c}{\vphantom{c}^{+}_{\downarrow}(\tau')}{X}\vphantom{X}^{\widetilde{1}\widetilde{4}}(\tau)=0,\]\vspace*{-7mm}
\[\rlcoupling{c}{\vphantom{c}^{+}_{\uparrow}(\tau')}{X}\vphantom{X}^{\widetilde{1}4}(\tau)
=g_{0}^{\widetilde{1}4}(\tau-\tau')(X^{\widetilde{1}\widetilde{1}}+X^{44})_{\tau'}\sin{\phi_{41}},\]\vspace*{-7mm}
\[\rlcoupling{c}{\vphantom{c}^{+}_{\downarrow}(\tau')}{X}\vphantom{X}^{\widetilde{1}4}(\tau)=0,\]\vspace*{-7mm}
\[\rlcoupling{c}{\vphantom{c}^{+}_{\uparrow}(\tau')}{X}\vphantom{X}^{1\widetilde{4}}(\tau)
=-g_{0}^{1\widetilde{4}}(\tau-\tau')(X^{11}+X^{\widetilde{4}\widetilde{4}})_{\tau'}\sin{\phi_{41}},\]\vspace*{-7mm}
\begin{equation}
\rlcoupling{c}{\vphantom{c}^{+}_{\downarrow}(\tau')}{X}\vphantom{X}^{1\widetilde{4}}(\tau)=0.
\end{equation}
Using (9), we have
\[\rlcoupling{c}{\vphantom{c}^{+}_{\uparrow}(\tau')}{c}\vphantom{c}_{\uparrow}(\tau)
=-g_{0}^{14}(\tau-\tau')(X^{11}+X^{44})_{\tau'}\cos^2{\phi_{41}}-\]\vspace*{-7mm}
\[-g_{0}^{\widetilde{1}\widetilde{4}}(\tau-\tau')(X^{\widetilde{1}\widetilde{1}}
+X^{\widetilde{4}\widetilde{4}})_{\tau'}\cos^2{\phi_{41}}-\]\vspace*{-7mm}
\[-g_{0}^{\widetilde{1}4}(\tau-\tau')(X^{\widetilde{1}\widetilde{1}}+X^{44})_{\tau'}\sin^2{\phi_{41}}-\]\vspace*{-7mm}
\begin{equation}
-g_{0}^{1\widetilde{4}}(\tau-\tau')(X^{11}+X^{\widetilde{4}\widetilde{4}})_{\tau'}\sin^2{\phi_{41}}.
\end{equation}

This result shows us that, in the case of the alloy analogy approximation
the pairing of Fermi-operators decomposes into the sum of terms that
are the projections on single-site states (because of
the action of $X^{rr}$ operators). This is the main difference from the case of ideal fermions,
where we have Green's functions as a result of the pairing.

Now, we can rewrite
\[\rlcoupling{c}{\vphantom{c}^{+}_{\uparrow}(\tau')}{c}\vphantom{c}_{\uparrow}(\tau)=\]\vspace*{-7mm}
\[-[g_{0}^{14}(\tau\!-\!\tau')\cos^2{\phi_{41}}\!+\!g_{0}^{1\widetilde{4}}(\tau\!-\!\tau')
\sin{\phi_{41}}]X^{11}(\tau')-\]\vspace*{-7mm}
\[-[g_{0}^{\widetilde{1}\widetilde{4}}(\tau\!-\!\tau')\cos^2{\phi_{41}}\!
+\!g_{0}^{\widetilde{1}4}(\tau\!-\!\tau')(\sin{\phi_{41}}]X^{\widetilde{1}\widetilde{1}}(\tau')-\]\vspace*{-7mm}
\[-[g_{0}^{14}(\tau\!-\!\tau')\cos^2{\phi_{41}}\!+\!g_{0}^{\widetilde{1}4}(\tau\!-\!\tau')
\sin{\phi_{41}}]X^{44}(\tau')-\]\vspace*{-7mm}
\[-[g_{0}^{\widetilde{1}\widetilde{4}}(\tau\!-\!\tau')\cos^2{\phi_{41}}\!+
\!g_{0}^{1\widetilde{4}4}(\tau\!-\!\tau')\sin{\phi_{41}}]X^{\widetilde{4}\widetilde{4}}(\tau')\equiv\]\vspace*{-7mm}
\[-g_{0\uparrow}^{(1)}X^{11}(\tau')-g_{0\uparrow}^{(\widetilde{1})}X^{\widetilde{1}\widetilde{1}}(\tau')
-g_{0\uparrow}^{(4)}X^{44}(\tau')-\]\vspace*{-7mm}
\begin{equation}
-g_{0\uparrow}^{(\widetilde{4})}X^{\widetilde{4}\widetilde{4}}(\tau')
=-\sum\limits_{r}g_{0\uparrow}^{(r)}X^{rr}(\tau').
\end{equation}

\section{Single-Site Green's Function}

In general, electron Green's function of the effective one-site
problem reads
\[\varUpsilon_{\sigma}(\tau-\tau')=-\frac{\langle Tc_{\sigma}
(\tau)c^{+}_{\sigma}(\tau')e^{-\beta H_{\rm eff}}\rangle}{\langle
e^{-\beta H_{\rm eff}}\rangle}=\]\vspace*{-7mm}
\begin{equation}
=-\frac{\langle Tc_{\sigma}(\tau)c^{+}_{\sigma}(\tau')
\widetilde{\sigma}(\beta)\rangle_0}{\langle\widetilde{\sigma}(\beta)
\rangle_0}.
\end{equation}

The numerator and denominator in this expression will be calculated
separately using an expansion in terms of the coherent potential
$J_{\sigma}(\tau-\tau')$. As the first step, we illustrate the
second order in this expansion with four operators of creation and
annihilation of electrons:
\[\langle T
c_{\uparrow}(\tau)c_{\uparrow}^{+}(\tau')c_{\uparrow}^{+}(\tau_1)c_{\uparrow}(\tau_2)\rangle_{0}=\]\vspace*{-7mm}
\[=\langle T
\rlcoupling{c}{\vphantom{c}^{+}_{\uparrow}(\tau)}{c}\vphantom{c}_{\uparrow}
(\tau)\rlcoupling{c}{\vphantom{c}^{+}_{\uparrow}(\tau_1)}{c}\vphantom{c}_{\uparrow}(\tau_2)
\rangle_{0}+\]\vspace*{-7mm}
\[+\langle T
\rlcoupling{c}{\vphantom{c}^{+}_{\uparrow}(\tau_1)}{c}\vphantom{c}_{\uparrow}
(\tau)\rlcoupling{c}{\vphantom{c}^{+}_{\uparrow}(\tau')}{c}\vphantom{c}_{\uparrow}(\tau_2)
\rangle_{0}=\]\vspace*{-7mm}
\[=-\sum\limits_{r}g_{0\uparrow}^{(r)}(\tau-\tau')g_{0\uparrow}^{(r)}(\tau_2-\tau_1)\langle
X^{rr}\rangle_0+\]\vspace*{-7mm}
\begin{equation}
+\sum\limits_{r}g_{0\uparrow}^{(r)}(\tau-\tau_1)g_{0\uparrow}^{(r)}(\tau_2-\tau')\langle
X^{rr}\rangle_0
\end{equation}
and
\[\langle T c_{\uparrow}(\tau)c_{\uparrow}^{+}(\tau')c_{\downarrow}^{+}(\tau_1)c_{\downarrow}(\tau_2)\rangle_{0}=\]\vspace*{-7mm}
\[=-\langle T \rlcoupling{c}{\vphantom{c}^{+}_{\uparrow}(\tau')}
{c}\vphantom{c}_{\uparrow}(\tau)\rlcoupling{c}
{\vphantom{c}^{+}_{\downarrow}(\tau_1)}{c}\vphantom{c}_{\downarrow}(\tau_2)
\rangle_{0}=\]\vspace*{-7mm}
\begin{equation}
=-\sum\limits_{r}g_{0\uparrow}^{(r)}(\tau-\tau')g_{0\downarrow}^{(r)}(\tau_2-\tau_1)\langle
X^{rr}\rangle_0.
\end{equation}
Here, the pairing of Fermi-operators is performed according to (25).
The diagonal $X$-operators, which appear during this procedure, we
multiply, by using the rule $X^{rr}X^{pp}=X^{rr}\delta_{rp}$. As a
result, only the averages $\langle X^{rr} \rangle_0$ are present.

We can also consider the third order in our expansion with six operators
of creation and annihilation of electrons:
\[\langle T c_{\uparrow}(\tau)c_{\uparrow}^{+}(\tau')c_{\uparrow}^{+}
(\tau_1)c_{\uparrow}(\tau_2)c_{\uparrow}^{+}(\tau_3)c_{\uparrow}(\tau_4)\rangle_{0}=\]\vspace*{-7mm}
\[=-\langle T \rlcoupling{c}{\vphantom{c}^{+}_{\uparrow}(\tau')}{c}
\vphantom{c}_{\uparrow}(\tau)c_{\uparrow}^{+}(\tau_1)c_{\uparrow}(\tau_2)c_{\uparrow}^{+}
(\tau_3)c_{\uparrow}(\tau_4) \rangle_{0}+\]\vspace*{-7mm}
\[+\langle T
c_{\uparrow}^{+}(\tau')\rlcoupling{c}{\vphantom{c}^{+}_{\uparrow}(\tau1)}
{c}\vphantom{c}_{\uparrow}(\tau)c_{\uparrow}(\tau_2)c_{\uparrow}^{+}(\tau_3)c_{\uparrow}(\tau_4)
\rangle_{0}+\]\vspace*{-7mm}
\[+\langle T
c_{\uparrow}^{+}(\tau')c_{\uparrow}^{+}(\tau_1)c_{\uparrow}(\tau_2)
\rangle_{0}
\rlcoupling{c}{\vphantom{c}^{+}_{\uparrow}(\tau_3)}{c}\vphantom{c}_{\uparrow}
(\tau)c_{\uparrow}(\tau_4)\rangle_0=\]\vspace*{-7mm}
\[=-\langle T
\rlcoupling{c}{\vphantom{c}^{+}_{\uparrow}(\tau')}{c}\vphantom{c}_{\uparrow}(\tau)
\rlcoupling{c}{\vphantom{c}^{+}_{\uparrow}(\tau_1)}{c}\vphantom{c}_{\uparrow}(\tau_2)
\rlcoupling{c}{\vphantom{c}^{+}_{\uparrow}(\tau_3)}{c}\vphantom{c}_{\uparrow}(\tau_4)\rangle_{0}+\]\vspace*{-7mm}
\[+\langle T
\rlcoupling{c}{\vphantom{c}^{+}_{\uparrow}(\tau')}{c}\vphantom{c}_{\uparrow}(\tau)
\rlcoupling{c}{\vphantom{c}^{+}_{\uparrow}(\tau_1)}{c}\vphantom{c}_{\uparrow}(\tau_4)
\rlcoupling{c}{\vphantom{c}^{+}_{\uparrow}(\tau_3)}{c}\vphantom{c}_{\uparrow}(\tau_2)\rangle_{0}+\]\vspace*{-7mm}
\[{c}\vphantom{c}_{\uparrow}(\tau_2)
\rlcoupling{c}{\vphantom{c}^{+}_{\uparrow}(\tau_1)}{c}\vphantom{c}_{\uparrow}(\tau)
\rlcoupling{c}{\vphantom{c}^{+}_{\uparrow}(\tau_3)}{c}\vphantom{c}_{\uparrow}(\tau_4)\rangle_{0}-\]\vspace*{-7mm}
\[-\langle T
\rlcoupling{c}{\vphantom{c}^{+}_{\uparrow}(\tau')}{c}\vphantom{c}_{\uparrow}(\tau_4)
\rlcoupling{c}{\vphantom{c}^{+}_{\uparrow}(\tau_1)}{c}\vphantom{c}_{\uparrow}(\tau)
\rlcoupling{c}{\vphantom{c}^{+}_{\uparrow}(\tau_3)}{c}\vphantom{c}_{\uparrow}(\tau_2)\rangle_{0}+\]\vspace*{-7mm}
\[+\langle T
\rlcoupling{c}{\vphantom{c}^{+}_{\uparrow}(\tau')}{c}\vphantom{c}_{\uparrow}(\tau_4)
\rlcoupling{c}{\vphantom{c}^{+}_{\uparrow}(\tau_1)}{c}\vphantom{c}_{\uparrow}(\tau_2)
\rlcoupling{c}{\vphantom{c}^{+}_{\uparrow}(\tau_3)}{c}\vphantom{c}_{\uparrow}(\tau)\rangle_{0}-\]
\begin{equation}
-\langle T
\rlcoupling{c}{\vphantom{c}^{+}_{\uparrow}(\tau')}{c}\vphantom{c}_{\uparrow}(\tau_2)
\rlcoupling{c}{\vphantom{c}^{+}_{\uparrow}(\tau_3)}{c}\vphantom{c}_{\uparrow}(\tau)
\rlcoupling{c}{\vphantom{c}^{+}_{\uparrow}(\tau_1)}{c}\vphantom{c}_{\uparrow}(\tau_4)\rangle_{0}.
\end{equation}
Finally,
\[\langle T
c_{\uparrow}(\tau)c_{\uparrow}^{+}(\tau')c_{\uparrow}^{+}(\tau_1)c_{\uparrow}
(\tau_2)c_{\uparrow}^{+}(\tau_3)c_{\uparrow}(\tau_4)\rangle_{0}=\]\vspace*{-5mm}
\[=\sum\limits_{r}g_{0\uparrow}^{(r)}(\tau-\tau')g_{0\uparrow}^{(r)}
(\tau_2-\tau_1)g_{0\uparrow}^{(r)}(\tau_4-\tau_3)\langle X^{rr}
\rangle_{0}-\]\vspace*{-5mm}
\[-\sum\limits_{r}g_{0\uparrow}^{(r)}(\tau-\tau')g_{0\uparrow}^{(r)}
(\tau_4-\tau_1)g_{0\uparrow}^{(r)}(\tau_2-\tau_3)\langle X^{rr}
\rangle_{0}-\]\vspace*{-5mm}
\[-\sum\limits_{r}g_{0\uparrow}^{(r)}(\tau_2-\tau')g_{0\uparrow}^{(r)}(\tau-\tau_1)
g_{0\uparrow}^{(r)}(\tau_4-\tau_3)\langle X^{rr}
\rangle_{0}+\]\vspace*{-5mm}
\[+\sum\limits_{r}g_{0\uparrow}^{(r)}(\tau_4-\tau')g_{0\uparrow}^{(r)}
(\tau-\tau_1)g_{0\uparrow}^{(r)}(\tau_2-\tau_3)\langle X^{rr}
\rangle_{0}-\]\vspace*{-5mm}
\[-\sum\limits_{r}g_{0\uparrow}^{(r)}(\tau_4-\tau')g_{0\uparrow}^{(r)}(\tau_2-\tau_1)
g_{0\uparrow}^{(r)}(\tau-\tau_3)\langle X^{rr}
\rangle_{0}+\]\vspace*{-5mm}
\begin{equation}
+\sum\limits_{r}g_{0\uparrow}^{(r)}(\tau_2-\tau')g_{0\uparrow}^{(r)}
(\tau-\tau_3)g_{0\uparrow}^{(r)}(\tau_4-\tau_1)\langle X^{rr}
\rangle_{0}.
\end{equation}

The similar procedure is also actual in the case of higher order
terms. Using the diagrammatic series, we can separate the connected and
disconnected ``vacuum'' (without external vertices) parts of
diagrams. The former form a geometric progression in the frequency
representation. The latter look like closed rings of different
lengths (created by unperturbed Green's function and coherent
potential lines) and give
exponential contributions in subspaces $|r\rangle$ after the summation of infinite series.

So, the numerator in Green's function $\langle T
c_{\uparrow}(\tau)c_{\uparrow}^{+}(\tau')\rangle_{\rm num}$ reads
\[\langle T c_{\uparrow}(\tau)c_{\uparrow}^{+}(\tau')\rangle_{\rm
num}=\]\vspace*{-5mm}
\[=\sum\limits_{r}\Big[g_{0\uparrow}^{(r)}(\omega_n)-g_{0\uparrow}^{(r)}(\omega_n)
J_{\uparrow}(\omega_n)g_{0\uparrow}^{(r)}(\omega_n)+\]\vspace*{-5mm}
\[+g_{0\uparrow}^{(r)}(\omega_n)J_{\uparrow}(\omega_n)g_{0\uparrow}^{(r)}
(\omega_n)J_{\uparrow}(\omega_n)g_{0\uparrow}^{(r)}(\omega_n)-\]\vspace*{-7mm}
\begin{equation}
-\dots\Big]\langle X^{rr}\rangle_0
e^{Q_{r}}=\sum\limits_{r}\frac{g_{0\uparrow}^{(r)}(\omega_n)}{1+g_{0\uparrow}^{(r)}(\omega_n)
J_{\uparrow}(\omega_n)}\langle X^{rr}\rangle_0 e^{Q_{r}}.
\end{equation}
Here, $Q_r$ in the analytical form is
\[Q_r=\sum\limits_{\omega_n}\sum\limits_{\sigma}g_{0\sigma}^{(r)}(\omega_n)J_{\sigma}(\omega_n)-\]\vspace*{-5mm}
\[-\frac{1}{2}\Big[\sum\limits_{\omega_n}\sum\limits_{\sigma}g_{0\sigma}^{(r)}(\omega_n)J_{\sigma}(\omega_n)\Big]^2+\]\vspace*{-5mm}
\[+\frac{1}{3}\Big[\sum\limits_{\omega_n}\sum\limits_{\sigma}g_{0\sigma}^{(r)}(\omega_n)J_{\sigma}
(\omega_n)\Big]^3-\dots=\]\vspace*{-5mm}
\begin{equation}
=\sum\limits_{\omega_n}\sum\limits_{\sigma}\ln(1+g_{0\sigma}^{(r)}(\omega_n)J_{\sigma}(\omega_n)).
\end{equation}
In particular,
\[Q_{1,\widetilde{1}}=\sum\limits_{\omega_n}\ln(1+g_{0\uparrow}^{(1,\widetilde{1})}(\omega_n)J_{\uparrow}(\omega_n))+\]\vspace*{-5mm}
\[+\sum\limits_{\omega_n}\ln(1+g_{0\downarrow}^{(1,\widetilde{1})}(\omega_n)J_{\downarrow}(\omega_n)),\]\vspace*{-5mm}
\[Q_{3,\widetilde{3}}=\sum\limits_{\omega_n}\ln(1+g_{0\downarrow}^{(3,\widetilde{3})}(\omega_n)J_{\downarrow}(\omega_n)),\]\vspace*{-5mm}
\begin{equation}
Q_{4,\widetilde{4}}=\sum\limits_{\omega_n}\ln(1+g_{0\uparrow}^{(4,\widetilde{4})}(\omega_n)J_{\uparrow}(\omega_n)).
\end{equation}
The next step is to calculate the denominator
\[\langle \widetilde{\sigma}(\beta)
\rangle_{0}=1-\int\limits_{0}^{\beta}d\tau_1\int\limits_{0}^{\beta}d\tau_2
\sum\limits_{\sigma}J_{\sigma}(\tau_1-\tau_2)\times\]\vspace*{-7mm}
\[\times\langle T c^{+}_{\sigma}(\tau_1)c_{\sigma}(\tau_2)\rangle_{0}+\frac{1}{2!}
\int\limits_{0}^{\beta}d\tau_1\dots\int\limits_{0}^{\beta}d\tau_4\times\]\vspace*{-7mm}
\[\times\sum\limits_{\sigma}\sum\limits_{\sigma'}J_{\sigma}(\tau_1-\tau_2)J_{\sigma'}(\tau_3-\tau_4)\times\]\vspace*{-7mm}
\begin{equation}
\times\langle T
c^{+}_{\sigma}(\tau_1)c_{\sigma}(\tau_2)c^{+}_{\sigma'}(\tau_3)c_{\sigma'}(\tau_4)\rangle_{0}-\dots
\end{equation}
In the diagrammatic representation, this series is expressed through the
set of ``vacuum'' diagrams. The final result could be expressed in
terms of contributions $Q_r$ of the above-mentioned ring diagrams.

\begin{figure}
\includegraphics[width=8cm]{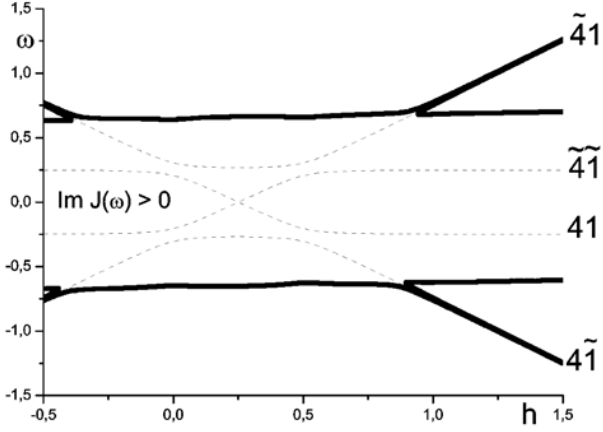}
\vskip-3mm\caption{ Dependence of electron band boundaries on the
asymmetry of the local anharmonic potential $h$ ($g=0.5$,
$\Omega=0.1$, $T=0.02$, $\mu=0$, $W=0.5$). Hereinafter, dashed lines
represent the energies of the  transitions between single-site
electron levels$ pq=\lambda_{pq}=\varepsilon_p-\varepsilon_q$
without hopping }
\end{figure}

In such a case, we have
\[\langle \sigma(\beta) \rangle_{0}=1+\sum\limits_{r}\Big[
Q_r+\frac{1}{2!} Q^2_r+\frac{1}{3!} Q^3_r+\]\vspace*{-5mm}
\begin{equation}
+\dots\Big]\langle X^{rr}\rangle_0=\sum\limits_{r} e^{Q_r}\langle
X^{rr}\rangle_0.
\end{equation}

Finally, our analytical result is
\begin{equation}
\langle T c^{+}_{\sigma}c_{\sigma}
\rangle=\frac{\sum\limits_{r}\frac{g_{0\sigma}^{(r)}(\omega_n)}{1+g_{0\sigma}^{(r)}(\omega_n)J_{\sigma}(\omega_n)}\langle
X^{rr}\rangle_0 e^{Q_{r}}}{\sum\limits_{p} e^{Q_p}\langle
X^{pp}\rangle_0},
\end{equation}
where $\sigma=\uparrow$ or $\downarrow$.

\section{Electron Energy Spectrum}

Now, we have the closed system of equations to calculate Green's
function $ G_{\uparrow}^{(a)}(\omega)$ and the coherent potential $
J_{\uparrow}(\omega)$:
\[G_{k}^{\sigma}(\omega)=\frac{1}{[\Xi_{\sigma}(\omega)]^{-1}-t_{k}},\]
\[G_{\sigma}^{(a)}(\omega)=\frac{1}{[\Xi_{\sigma}(\omega)]^{-1}-J_{\sigma}
(\omega)}=\frac{1}{N}\sum\limits_{k}G_k^{\sigma}(\omega),\]\vspace*{-5mm}
\[G_{\sigma}^{(a)}(\omega)=\frac{\sum\limits_{r}\frac{g_{0\sigma}^{(r)}
(\omega_n)}{1+g_{0\sigma}^{(r)}(\omega_n)J_{\sigma}(\omega_n)}\langle
X^{rr} \rangle_0 e^{Q_{r}}}{\sum\limits_{p} e^{Q_p}\langle
X^{pp}\rangle_0},\]\vspace*{-5mm}
\begin{equation}
Q_r=\sum\limits_{\omega_n}\sum\limits_{\sigma}\ln(1+g_{0\sigma}^{(r)}(\omega_n)J_{\sigma}(\omega_n)).
\end{equation}

\begin{figure}
\includegraphics[width=8cm]{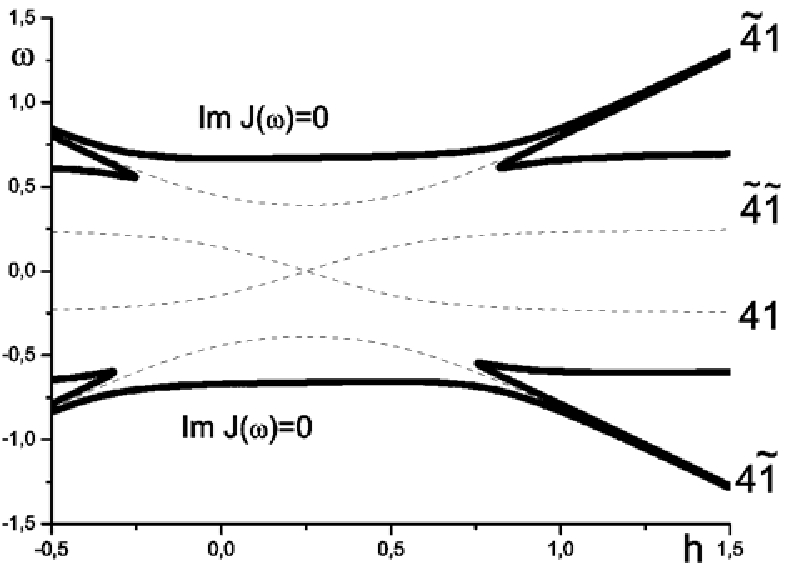}
\vskip-3mm\caption{ Dependence of electron band boundaries on the
asymmetry of the local anharmonic potential $h$ ($g=0.5$,
$\Omega=0.3$, $T=0.02$, $\mu=0$, $W=0.5$) }\vskip3mm
\end{figure}

\begin{figure}
\includegraphics[width=8cm]{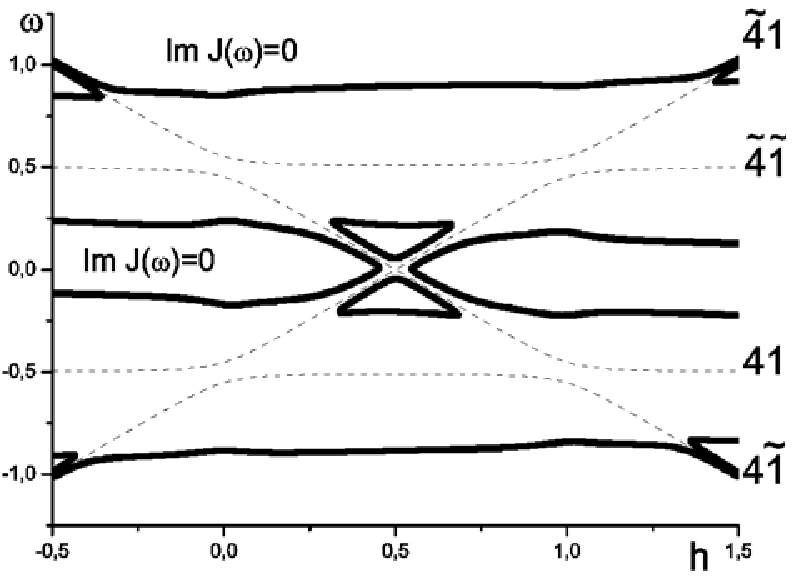}
\vskip-3mm\caption{ Dependence of electron band boundaries on the
asymmetry of the local anharmonic potential $h$ ($g=1.0$,
$\Omega=0.1$, $T=0.02$, $\mu=0$, $W=0.5$) }
\end{figure}

Here, to calculate $\langle X^{rr}\rangle_0,$ we use the iterative
process: the next $\langle X^{rr}\rangle_0$ value depends on the
previous $\langle X^{rr}\rangle_0^{'}$ one as
\[\langle X^{rr}\rangle_0=\frac{\langle
X^{rr}\rangle_0^{'}e^{Q_{r}}}{\sum\limits_{p=1,\widetilde{1},4,\widetilde{4}}\langle
X^{pp}\rangle_0 e^{Q_{p}}}.\]

The initial averages are taken from the Boltzmann distribution
$\langle X^{rr}\rangle_0=\frac{e^{-\beta\varepsilon_r}}{\sum\limits_p
e^{-\beta\varepsilon_p}}$.

To sum over ${\bf k},$  we use the semielliptic density of
states $\rho_0(t)=\frac{2}{\pi W^2}\sqrt{W^2-t^2}$. In this case,
$J_{\sigma}(\omega)=\frac{W^2}{4}G^{(a)}_{\sigma}(\omega)$
\cite{Voll}, and our final equation for the coherent potential $
J_{\sigma}(\omega)$ is as follows:
\begin{equation}
J_{\sigma}(\omega_n)=\frac{W^2}{4}\frac{\sum\limits_{r}\frac{g_{0\sigma}^{(r)}(\omega_n)}
{1+g_{0\sigma}^{(r)}(\omega_n)J_{\sigma}(\omega_n)}\langle
X^{rr}\rangle_0 e^{Q_{r}}} {\sum\limits_{p} e^{Q_p}\langle
X^{pp}\rangle_0}.
\end{equation}

In a usual way, we perform the analytical continuation on the real axis
$(i\omega_n \rightarrow \omega - i\delta),$ and only the solutions with
$\Im{J_{\sigma}(\omega)}>0$ must be considered.

Electron band boundaries are determined from the condition
$\Im{J_{\sigma}(\omega)}\rightarrow 0$. Their dependences on
the asymmetry of the local anharmonic potential are shown in Figs. 1--4.

\begin{figure}
\includegraphics[width=8cm]{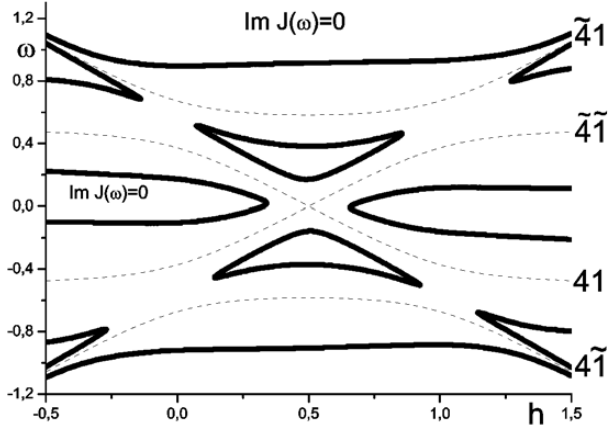}
\vskip-3mm\caption{ Dependence of electron band boundaries on the
asymmetry of the local anharmonic potential $h$ ($g=1.0$,
$\Omega=0.3$, $T=0.02$, $\mu=0$, $W=0.5$) }\vskip3mm
\end{figure}

\begin{figure}
\includegraphics[width=7.8cm]{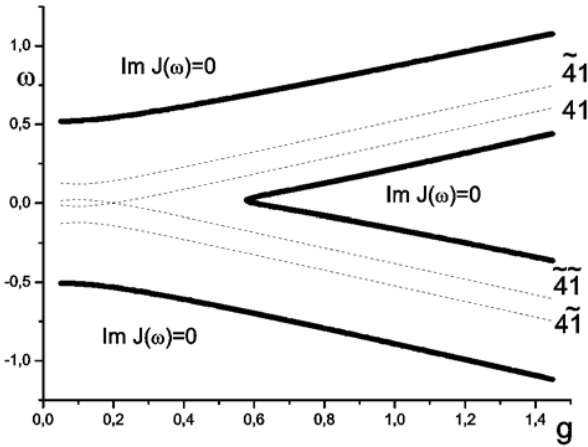}
\vskip-3mm\caption{ Dependence of electron band boundaries on the
pseudospin-electron interaction constant $g$ ($h=0.1$, $\Omega=0.1$,
$T=0.02$, $\mu=0$, $W=0.5$) }
\end{figure}

One can see the effect of the tunneling-like level splitting $\Omega$
on the width of existing bands (we can compare the plots with constant
$g=0.5$ and to see that the increase of $\Omega $ from 0.1 to 0.3
leads to a noticable broadening of bands).

\begin{figure}
\includegraphics[width=7.9cm]{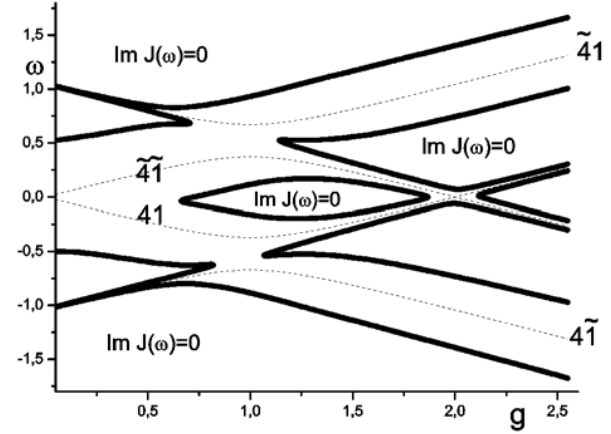}
\vskip-3mm\caption{ Dependence of electron band boundaries on the
pseudospin-electron interaction constant $g$ ($h=1.0$, $\Omega=0.3$,
$T=0.02$, $\mu=0$, $W=0.5$) }\vskip2mm
\end{figure}
\begin{figure}
\includegraphics[width=8cm]{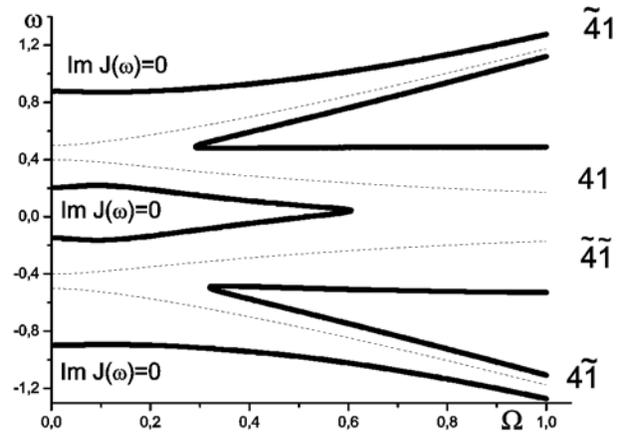}
\vskip-2mm\caption{ Dependence of electron band boundaries on the
tunneling-like level splitting $\Omega$ ($g=1.0$, $h=0.1$, $T=0.02$,
$\mu=0$, $W=0.5$) }\vskip-1mm
\end{figure}

\begin{figure*}
\includegraphics[width=7.0cm]{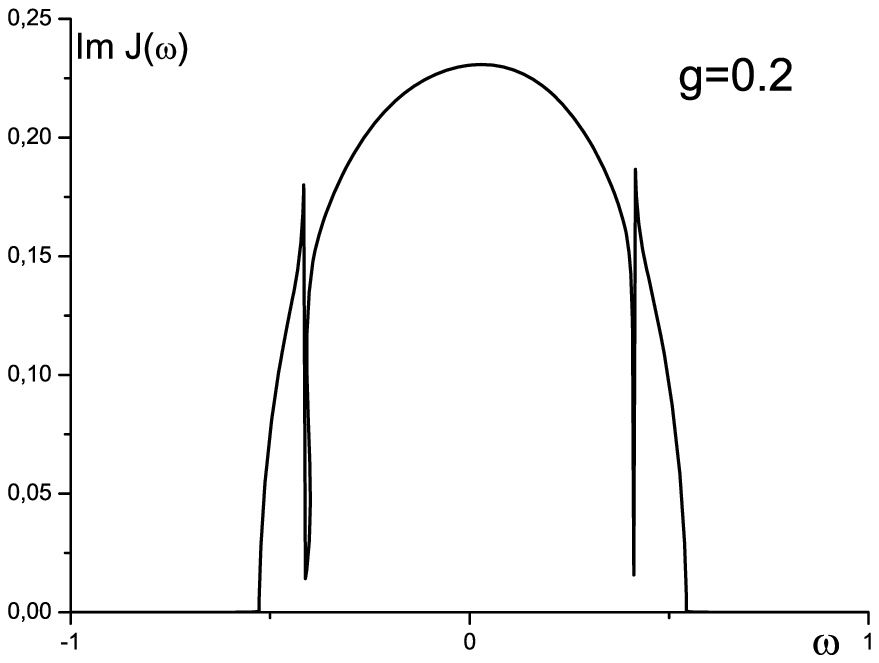}\hspace{0.5cm}\includegraphics[width=7.0cm]{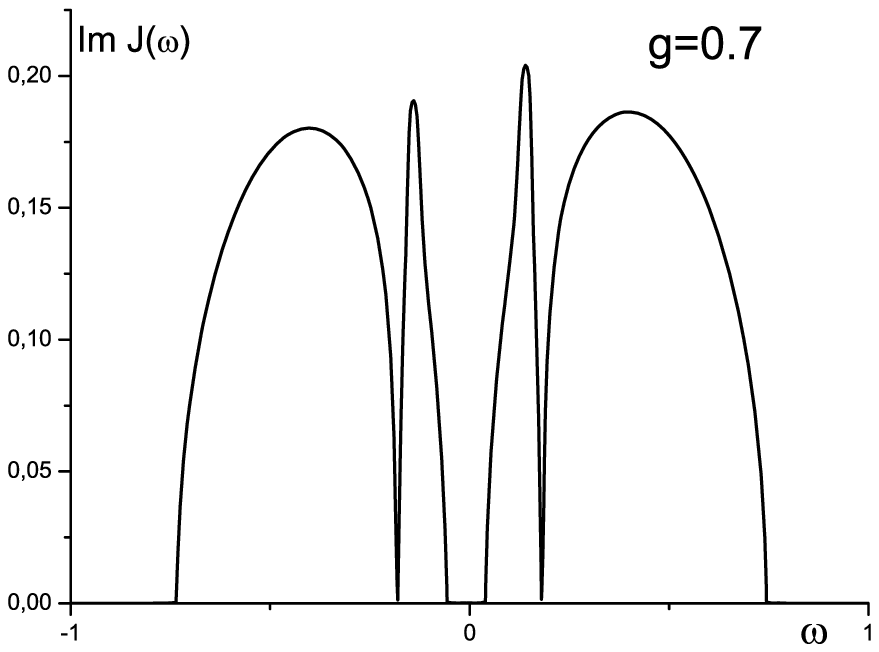}\\[2mm]
\includegraphics[width=7.0cm]{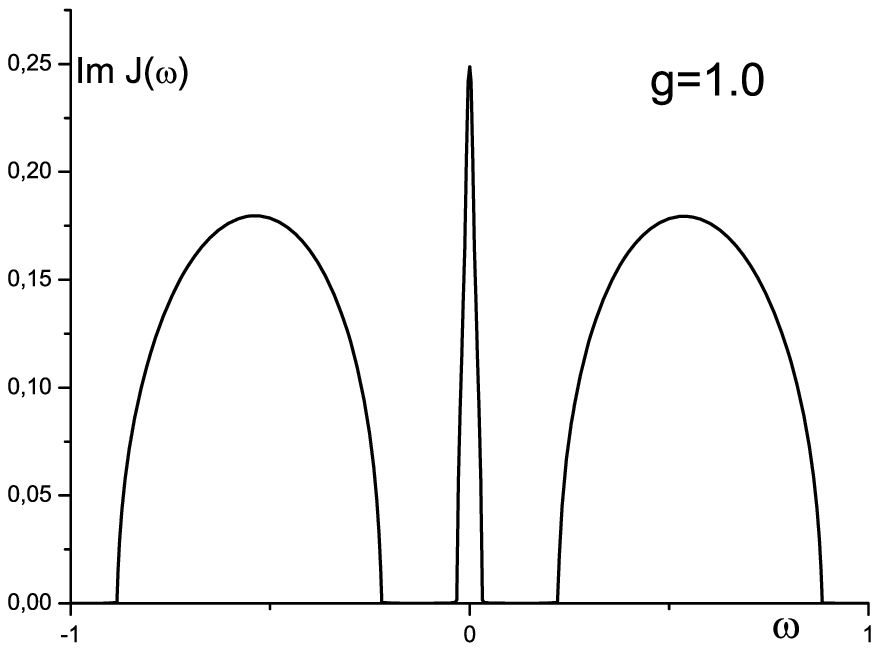}\hspace{0.5cm}\includegraphics[width=7.0cm]{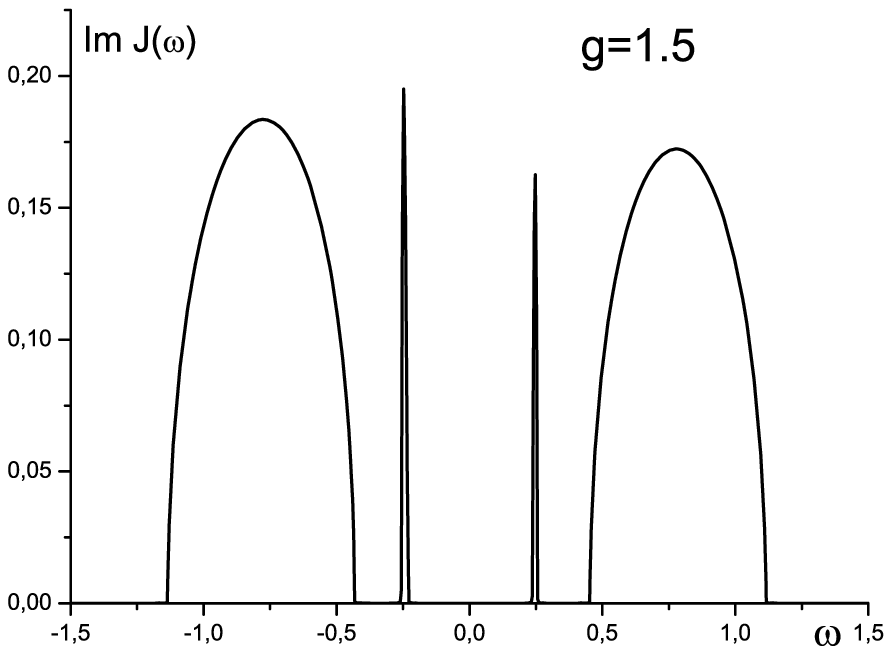}
\vskip-3mm\caption{Electron density of states at different values of
the pseudospin-electron interaction constant $g$ ($T=0.02$, $\mu=0$,
$W=0.5$) }
\end{figure*}

One can see also how the pseudospin-electron interaction $g$ leads
to the appearance of two additional bands, when $h \approx
\frac{g}{2},$ and one band otherwise (this can be seen with
comparing the plots with different $g,$ but with the same $\Omega$).
The dependences of electron band boundaries on the
pseudospin-electron interaction constant $g$ are shown in
Figs.~5--6. One can see that there exists some critical $g \approx
W,$ at which the additional gap appears in the spectrum. This result
is similar to the Mott transition type reconstruction obtained in
\cite{JPS} for the pseudospin-electron model without tunneling-like
level \mbox{splitting.}

Similarly to the dependence of the local anharmonic potential on the
asymmetry, we can see also how an increase of $\Omega$ leads to the
band width growth. At $g \approx 2h,$ the additional bands appear in
the \mbox{spectrum.}

The dependences of electron band boundaries on the tunneling-like
level splitting $\Omega$ are shown in Fig.~7. Here, one can see the
presence of two critical values $\Omega\approx 0.3$ (band splitting)
and $\Omega\approx 0.6$ (band \mbox{emerging).}

The densities of states $\rho_{\sigma}(\omega)=\frac{1}{\pi} {\rm
Im}\, G_{\sigma}^{(a)} \omega-i0^+) $ for different values of the
pseudospin-electron interaction constant $g$ are shown in Fig. 8.
Initially, we observe the splitting of one band at $g\approx W$.
Afterwards at $g\approx 2h,$ two central bands emerge, and the
further increase of $g$ gives four bands in the electron spectrum.

\section{Conclusions}

1)  The electron energy spectrum of the pseudospin-electron model is
considered. For this purpose, the dynamical mean field method is
applied. The effective single-site problem is solved within an original
approach based on the use of a generalization of Wick's theorem. The
alloy analogy approximation is used.

\noindent 2) Electron transitions with a possibility of the simultaneous
flip-over of the pseudospin (or at its unchanged orientation)
manifest themselves in a complication of the electron spectrum
which consists in the appearance of additional subbands. The effects of
bond splitting and creation of new gaps take place, depending on
the longitudinal field $h$, interaction constant $g,$ and transverse
field $\Omega$ values (Figs.~1--7).

\noindent 3) The rearrangement of the spectrum, similarly to the Mott
transition, occurs not only due to the on-site peudospin-electron
interaction, but also under influence of the transverse field that
intertangles the states with different pseudospin orientations.

\noindent 4) The obtained results show that, in the real system,
which exhibits the local anharmonicity of lattice vibrations, the
metal-insulator transition due to the short-ranged electron
correlation is influenced by the anharmonic subsystem. Changing the
parameters of local anharmonicity (e.g., the potential
well shape), one can affect the conditions of the appearance of a a gap.

\noindent 5) The obtained results (Fig.~7) point out, in particular,
that the additional gaps can appear in the fermion band spectrum  at the
increase of $\Omega$. We have considered the case where, at $\Omega=0,$
the spectrum is already split. The problem in such a limit is
reduced to the Falikov--Kimball model with a Mott gap at the chosen
values of model parameters. In the case where PEM is applied to the
boson-fermion mixtures on a lattice, the increase of the $\Omega$
parameter corresponds to the deepening into the phase with BE-condensate
$(\Omega \sim \sum t^{b}_{ij}\langle S^x_j\rangle)$. Additional gaps
could appear at certain critical values of the order parameter
$\langle S^x\rangle$. This problem (which now attracts an attention
\cite{21}) requires, however, a more detailed investigation. In the
present study, based on the standard version of PEM, the boson
transfer is taken into account in the mean field approximation. At
the same time, the contributions of the collective pseudospin
dynamics to the fermion spectrum can be essential.

\noindent 6) It should be mentioned that we use an approach based
on the formalism of $X$-operators. Such a scheme gives a
possibility to consider, in a unified way, the creation of
composite excitations (fermion+boson; fermion+boson hole) and
their contributions to the total spectral density.

\noindent 7) The more complete analysis of a reconstruction of the energy
spectrum will be a subject of our subsequent consideration. It is
referred, in particular, to the region of half-filling ($\mu\sim 0$,
$h\sim\frac{g}{2}$), where the
instability with respect to the appearance of a modulated (CDW-like)
phase takes place in PEM at low enough temperatures. In such a case, the additional complication of
the spectrum will arise as a result.

\vspace*{-5mm}
\rezume{І.В. Стасюк, В.О. Краснов}{%
ЕНЕРГЕТИЧНИЙ СПЕКТР\\ ПСЕВДОСПІН-ЕЛЕКТРОННОЇ МОДЕЛІ\\ В МЕТОДІ
ДИНАМІЧНОГО СЕРЕДНЬОГО ПОЛЯ} {Досліджено псевдоспін-електронну
модель у випадку нескінченної взаємодії відштовхування електронів на
вузлі. Електронний енергетичний спектр моделі розраховано в рамках
методу динамічного середнього поля (ДСП) та наближення сплаву.
Досліджено вплив псевдоспін-електронної взаємодії, локального поля
асиметрії та тунельного розщеплення рівнів на існування та кількість
електронних підзон.  Обговорено можливість застосування
псевдоспін-електронної моделі  до розрахунку енергетичного спектра
бозон-ферміонних сумішей в оптичних ґратках.}

\end{document}